\documentclass{rnaastex}
\usepackage{textcomp}

\usepackage{lineno}

\shorttitle{Proton modulation PAMELA}
\shortauthors{M. Martucci et al.}

\begin{document}

\title{Proton fluxes measured by the PAMELA experiment from the minimum to the maximum solar
activity for the 24th solar cycle}

\correspondingauthor{Matteo Martucci}
\email{mmartucci@roma2.infn.it}

\author{M.~Martucci}
\affiliation{University of Rome ``Tor Vergata'', Department of Physics,  I-00133 Rome, Italy}
\affiliation{INFN, Laboratori Nazionali di Frascati, Via Enrico Fermi 40, I-00044 Frascati, Italy}

\author{R. Munini}
\affil{INFN, Sezione di Trieste I-34149 Trieste, Italy}

\author{M.~Boezio}
\affiliation{INFN, Sezione di Trieste I-34149 Trieste, Italy}

\author{V.~Di~Felice}
\affiliation{INFN, Sezione di Rome ``Tor Vergata'', I-00133 Rome, Italy}
\affiliation{Space Science Data Center - Agenzia Spaziale Italiana, via del Politecnico, s.n.c., I-00133, Roma, Ital}

\author{O.~Adriani}
\affiliation{University of Florence, Department of Physics, I-50019 Sesto Fiorentino, Florence, Italy}
\affiliation{INFN, Sezione di Florence, I-50019 Sesto Fiorentino, Florence, Italy}

\author{G.~C.~Barbarino}
\affiliation{University of Naples ``Federico II'', Department of Physics, I-80126 Naples, Italy}
\affiliation{INFN, Sezione di Naples,  I-80126 Naples, Italy}

\author{G.~A.~Bazilevskaya}
\affiliation{Lebedev Physical Institute, RU-119991, Moscow, Russia}

\author{R.~Bellotti}
\affiliation{University of Bari, Department of Physics, I-70126 Bari, Italy}
\affiliation{INFN, Sezione di Bari, I-70126 Bari, Italy}

\author{M.~Bongi}
\affiliation{University of Florence, Department of Physics, I-50019 Sesto Fiorentino, Florence, Italy}
\affiliation{INFN, Sezione di Florence, I-50019 Sesto Fiorentino, Florence, Italy}

\author{V.~Bonvicini}
\affiliation{INFN, Sezione di Trieste I-34149 Trieste, Italy}

\author{S.~Bottai}
\affiliation{INFN, Sezione di Florence, I-50019 Sesto Fiorentino, Florence, Italy}

\author{A.~Bruno}
\affiliation{INFN, Sezione di Bari, I-70126 Bari, Italy}

\author{F.~Cafagna}
\affiliation{University of Bari, Department of Physics, I-70126 Bari, Italy}
\affiliation{INFN, Sezione di Bari, I-70126 Bari, Italy}

\author{D.~Campana}
\affiliation{INFN, Sezione di Naples,  I-80126 Naples, Italy}

\author{P.~Carlson}
\affiliation{KTH, Department of Physics, Oskar Klein Centre for Cosmoparticle Physics, AlbaNova University Centre, SE-10691 Stockholm, Sweden}

\author{M.~Casolino}
\affiliation{INFN, Sezione di Rome ``Tor Vergata'', I-00133 Rome, Italy}
\affiliation{RIKEN, Advanced Science Institute, Wako-shi, Saitama, Japan}

\author{G.~Castellini}
\affiliation{IFAC, I-50019 Sesto Fiorentino, Florence, Italy}

\author{C.~De~Santis}
\affiliation{INFN, Sezione di Rome ``Tor Vergata'', I-00133 Rome, Italy}

\author{A.~M.~Galper}
\affiliation{National Research Nuclear University MEPhI, RU-115409 Moscow}

\author{A.~V.~Karelin}
\affiliation{National Research Nuclear University MEPhI, RU-115409 Moscow}

\author{S.~V.~Koldashov}
\affiliation{National Research Nuclear University MEPhI, RU-115409 Moscow}

\author{S.~Koldobskiy}
\affiliation{National Research Nuclear University MEPhI, RU-115409 Moscow}

\author{S.~Y.~Krutkov}
\affiliation{Ioffe Physical Technical Institute,  RU-194021 St. Petersburg, Russia}

\author{A.~N.~Kvashnin}
\affiliation{Lebedev Physical Institute, RU-119991, Moscow, Russia}

\author{A.~Leonov}
\affiliation{National Research Nuclear University MEPhI, RU-115409 Moscow}

\author{V.~Malakhov}
\affiliation{National Research Nuclear University MEPhI, RU-115409 Moscow}

\author{L.~Marcelli}
\affiliation{INFN, Sezione di Rome ``Tor Vergata'', I-00133 Rome, Italy}

\author{N.~Marcelli}
\affiliation{INFN, Sezione di Rome ``Tor Vergata'', I-00133 Rome, Italy}

\author{A.~G.~Mayorov}
\affiliation{National Research Nuclear University MEPhI, RU-115409 Moscow}

\author{W.~Menn}
\affiliation{Universit\"{a}t Siegen, Department of Physics, D-57068 Siegen, Germany}

\author{M. Merge'}
\affiliation{INFN, Sezione di Rome ``Tor Vergata'', I-00133 Rome, Italy}
\affiliation{University of Rome ``Tor Vergata'', Department of Physics,  I-00133 Rome, Italy}

\author{V.~V.~Mikhailov}
\affiliation{National Research Nuclear University MEPhI, RU-115409 Moscow}

\author{E.~Mocchiutti}
\affil{INFN, Sezione di Trieste I-34149 Trieste, Italy}

\author{A.~Monaco}
\affiliation{University of Bari, Department of Physics, I-70126 Bari, Italy}
\affiliation{INFN, Sezione di Bari, I-70126 Bari, Italy}

\author{N.~Mori}
\affiliation{INFN, Sezione di Florence, I-50019 Sesto Fiorentino, Florence, Italy}

\author{G.~Osteria}
\affiliation{INFN, Sezione di Naples,  I-80126 Naples, Italy}

\author{B.~Panico}
\affiliation{INFN, Sezione di Naples,  I-80126 Naples, Italy}

\author{P.~Papini}
\affiliation{INFN, Sezione di Florence, I-50019 Sesto Fiorentino, Florence, Italy}

\author{M.~Pearce}
\affiliation{KTH, Department of Physics, Oskar Klein Centre for Cosmoparticle Physics, AlbaNova University Centre, SE-10691 Stockholm, Sweden}

\author{P.~Picozza}
\affiliation{INFN, Sezione di Rome ``Tor Vergata'', I-00133 Rome, Italy}
\affiliation{University of Rome ``Tor Vergata'', Department of Physics,  I-00133 Rome, Italy}

\author{M.~Ricci}
\affiliation{INFN, Laboratori Nazionali di Frascati, Via Enrico Fermi 40, I-00044 Frascati, Italy}

\author{S.~B.~Ricciarini}
\affiliation{IFAC, I-50019 Sesto Fiorentino, Florence, Italy}

\author{M.~Simon}
\affiliation{Universit\"{a}t Siegen, Department of Physics, D-57068 Siegen, Germany}

\author{R.~Sparvoli}
\affiliation{INFN, Sezione di Rome ``Tor Vergata'', I-00133 Rome, Italy}
\affiliation{University of Rome ``Tor Vergata'', Department of Physics,  I-00133 Rome, Italy}

\author{P.~Spillantini}
\affiliation{University of Florence, Department of Physics, I-50019 Sesto Fiorentino, Florence, Italy}
\affiliation{INFN, Sezione di Florence, I-50019 Sesto Fiorentino, Florence, Italy}

\author{Y.~I.~Stozhkov}
\affiliation{Lebedev Physical Institute, RU-119991, Moscow, Russia}

\author{A.~Vacchi}
\affiliation{INFN, Sezione di Trieste I-34149 Trieste, Italy}
\affiliation{University of Udine, Department of Mathematics and Informatics, I-33100 Udine, Italy}

\author{E.~Vannuccini}
\affiliation{INFN, Sezione di Florence, I-50019 Sesto Fiorentino, Florence, Italy}

\author{G.~Vasilyev}
\affiliation{Ioffe Physical Technical Institute,  RU-194021 St. Petersburg, Russia}

\author{S.~A.~Voronov}
\affiliation{National Research Nuclear University MEPhI, RU-115409 Moscow}

\author{Y.~T.~Yurkin}
\affiliation{National Research Nuclear University MEPhI, RU-115409 Moscow}

\author{G.~Zampa}
\affiliation{INFN, Sezione di Trieste I-34149 Trieste, Italy}

\author{N.~Zampa}
\affiliation{INFN, Sezione di Trieste I-34149 Trieste, Italy}

\author{M. S. ~Potgieter}
\affiliation{North-West University, Centre for Space Research, 2520 Potchefstroom, South Africa}

\author{J. L. ~Raath}
\affiliation{North-West University, Centre for Space Research, 2520 Potchefstroom, South Africa}

\begin{abstract}
Precise measurements of the time-dependent intensity of the low energy ($<50$ GeV) galactic cosmic rays are fundamental to test and 
improve the models which describe their propagation inside the heliosphere. Especially, data spanning different solar activity periods, i.e. from minimum to maximum, are needed to achieve comprehensive understanding of such physical phenomenon. 
The minimum phase between the 23$^{rd}$ and the 24$^{th}$ solar cycles was peculiarly long, extending up to the beginning of 2010 and followed by the maximum phase, reached during early 2014. 
In this paper, we present proton differential spectra measured from January 2010 to February 2014 by the PAMELA experiment. 
For the first time the galactic cosmic ray proton intensity was  studied over a wide energy range (0.08-50 GeV) 
by a single apparatus from a minimum to a maximum period of solar activity. 
The large statistics allowed the time variation to be investigated on a nearly monthly basis.
Data were compared and interpreted in the context of a state-of-the-art three-dimensional model describing the galactic cosmic rays propagation through the heliosphere. 
\end{abstract}


\vspace{2cm}

\section{Introduction}

The energy spectra of galactic cosmic rays (GCRs), measured at Earth, are significantly influenced by the Sun's activity. Traversing the heliosphere, GCRs interact with the expanding 
solar wind and its embedded turbulent magnetic field, undergoing convection, diffusion, adiabatic energy losses and particle drifts because of the global curvature and gradients 
of the heliospheric magnetic field. As a consequence, 
the intensity of GCRs at Earth decreases with respect to the GCR energy spectrum outside the heliosphere, Local Interstellar Spectrum (LIS). 
This solar modulation has large effects on low energy cosmic rays (less than
a few GeVs), while the effects gradually subside as the energy increases, becoming negligible above a few tens of GeV (e.g. \cite{Strauss_2014a}).  
This modulation mechanism depends on the particle species, their charge and energy per nucleon (or rigidity) and it changes with time, determined by solar activity e.g.
following the 11-year cycle and the 22-year magnetic polarity cycle (see also e.g. \cite{Potgieter_2013}).
Precise measurements of GCR spectra, at different phases of the solar cycle, are essential to understand the various processes affecting the propagation of cosmic
rays in the heliosphere (e.g. see \cite{nuovocimento,AMS02}). 

Since the 1950s the variability of the galactic cosmic-ray flux has been constantly monitored by a network of ground-based neutron monitors 
(e.g. see \cite{Moraal_2000,Shea_Smart_2000,Usoskin_2005}). However, the intensity of the galactic cosmic rays was indirectly inferred by 
these detectors measuring the nucleons produced by the nuclear cascade  generated by cosmic rays interacting with the atmosphere. The PAMELA
(Payload for Antimatter/Matter Exploration and Light-nuclei Astrophysics) space-borne experiment~\citep{Picozza_2007, Boezio_2009} provided direct
measurements of the cosmic ray energy spectra and composition. The apparatus collected data from July 2006 to January 2016, covering the most recent
solar activity period, between cycle 23$^{rd}$ and the current cycle 24$^{th}$. 
Measurements of the proton differential  energy spectra  provided by the PAMELA instrument during the most recent solar minimum (from mid-2006 to the end of 2009) 
in the energy range from $80$~MeV to $50$~GeV,  have already been published~\citep{Adriani_2013}.  The evolution of the low energy galactic proton spectra on 
a solar-rotation-time basis (Carrington rotation\footnote{Mean synodic rotational period of the Sun surface, corresponding to about 27.28 days, see 
\citet{Carrington_1863}.})  was presented, providing the first measurements of the changing solar modulation over a wide energy range and for a very quiet solar minimum.
This period showed an extraordinary quiet heliosphere and unusually prolonged minimum. It was expected that the new solar cycle would begin early in 2008, instead minimum modulation conditions continued until the end of 2009. 
As a consequence, the highest low energy proton intensities since the beginning of the space age was registered during December 2009 
(e.g. see \cite{Mewaldt_2010}), which was unexpected given the solar
magnetic field polarity epoch at that time (e.g. see \cite{Potgieter_2013_Braz, Strauss_2014}). 
Results of a state-of-the-art full three dimensional (3D) model \citep{Potgieter_2014} was used to reproduce the PAMELA observational data.
 This model was based on the solution of the Parker transport equation, taking into account  all the physical processes involved in solar modulation and simulating the 
 solar minimum conditions of the 23/24 cycles~\citep{Potgieter_2014, Vos_2015}.
Similar studies were conducted on the effects of solar modulation on cosmic ray electrons \citep{Adriani_el_2015,Potgieter_el_2015} from which the dependence 
of the solar modulation on a particle's charge sign was observed \citep{e_p_2017}.
Following this extraordinarily deep minimum, the subsequent increase in the solar activity appeared remarkably weak in terms of e.g. sunspot number, solar wind speed and
number of solar events~\citep{Schroder_2017,Aslam_2015,Jiang_2015}. According to various solar activity data, the maximum of cycle 24$^{th}$ occurred in early 2014, with 
an estimated changing in the global axial dipole sign taking place in October 2013, while northern and southern polar fields reversing in November 2012 and March 2014, 
respectively~\citep{Sun_2015}.

The results presented in this paper refer to the evolution of the proton intensity from the end of the last solar minimum (January 2010) until the maximum of cycle 24$^{th}$ (February 2014). The evolution of the  low energy proton spectrum was studied on a solar rotation period basis, similarly to the previous publication~\citep{Adriani_2013}.

In the following, a brief description of the mission and details about the data analysis will be presented. 
Results on the proton flux measurements from 2010 to 2014 in the energy range from 80 MeV - 50 GeV are then presented, compared with the results of the mentioned 3D numerical model simulating the same heliospheric condition of the data-taking period, and discussed in the 
framework of a solar modulation theory. 

\section{Instrument and data analysis}\label{Sec:analysis}

After its launch, on June $15$, 2006 the PAMELA experiment had been almost continuously taking data until January 2016. 
The experiment was located on board the Resurs-DK1 Russian satellite placed by a Soyuz rocket at a highly inclined ($70^{\circ}$) elliptical orbit between 
$350$ km and  $600$ km height, changed into a circular one of $580$ km in September 2010.
The satellite quasi-polar orbit allowed the PAMELA instrument to sample low cutoff-rigidity orbital regions for a considerable amount of time, making it suitable for low energy particle studies. 
The apparatus consisted of a combination of detectors that provided 
information for particle identification and precise energy measurements.
These detectors were
(from top to bottom): a Time-of-Flight system;
a magnetic spectrometer; an anti-coincidence system;
an electromagnetic imaging calorimeter; a shower tail
catcher scintillator and a neutron detector. 
Detailed information about the instrument can be found in \citep{Picozza_2007,Adriani_2014_physrep,nuovocimento}. 

The proton fluxes were evaluated on a Carrington rotation basis according to the official  listing (\url{http://umtof.umd.edu/pm/crn/}). No isotopic separation (proton/deuterium) was performed in this analysis.
The present analysis spans the period between Carrington rotation 2092 and 2146 (January 2010 - February 2014). 
Most of these observations took place during a high solar activity period characterized by numerous solar events even if it should be recalled that
solar cycle 24 is considerably less active than the three preceding cycles (e.g. see \cite{Schroder_2017}). A significant fraction of these events 
produced high energy particles (mostly protons and Helium nuclei with energies up to a few GeV). This particles reached the Earth orbit 
and were indistinguishable from the GCR component collected by the PAMELA instrument. For a proper study of the solar modulation of GCRs
this solar component 
had to be excluded. The approach used in this work was to remove the periods in which this contamination was present. 
Data were excluded for the duration of the solar event using the information recorded 
by the low energy ($>60$ MeV) proton channel of GOES-15 (\url{ftp://satdat.ngdc.noaa.gov/sem/goes/data/}). 
Solar events have been studied by the PAMELA experiment and have been the topic of other publications (e.g. see \cite{SEP}).
Also the periods of Forbush decreases\footnote{A decrease of the galactic cosmic ray intensity observed in the Earth vicinity over a period of several 
days caused by transient solar phenomena such as interplanetary coronal mass ejections.} observed by the PAMELA instrument (e.g. \cite{Munini_2017})  were excluded from the analysis. 

The analysis procedure used in this work was similar to the one applied to the proton data over the solar minimum period presented and discussed in \cite{Adriani_2013}.
The fluxes were evaluated as follows:

\begin{equation}
  \phi(E) = \frac{N(E)}{\epsilon(E) \times G(E) \times T \times
    \Delta E} 
\end{equation}
where $N(E)$ is the unfolded count distribution, $\epsilon(E)$ the
efficiencies of the particle selections, $G(E)$ the 
geometrical factor, $T$ the live-time and $\Delta E$ the width of  
the energy interval. 

The large proton statistics permitted the study of the selection efficiencies in-flight for each Carrington rotation. 
This was particularly relevant for the time dependent track reconstruction efficiency, which varied from $\sim$20$\%$ in December 2009 to $\sim$15$\%$ at the beginning of 2014, 
because of the sudden failure of some front-end chips of the tracking system (see  \cite{Adriani_el_2015}).
This experimental information was combined with Monte Carlo simulation (performed with GEANT4 \citep{Agostinelli_2003}) to properly
reproduce the in-flight setup 
configuration as describe in \citep{Adriani_2011,Adriani_el_2015}.

The geometrical factor, i.e. the requirement of triggering and
containment, at least 1.5~mm away from the magnet
walls and the TOF-scintillator edges,  
was estimated with the full simulation 
of the apparatus and was found 
to be
constant at 19.9 cm$^{2}$~sr.
The live time was provided by an on-board clock that
timed the periods during which the apparatus was waiting for a
trigger. 

Both the response of the spectrometer (i.e. the rigidity resolution) and the ionization energy losses suffered by the protons crossing the detector caused 
a migration of proton events from one energy bin to another. 
To account for these effects and obtain the unfolded count distribution a Bayesian unfolding procedure, as
described in \cite{DAgostini_1995}, was applied (see also \cite{Adriani_el_2015} and \cite{Munini_2015}).
The detector response matrix was obtained from the simulation and calculated over each Carrington rotation to follow any change in the instrumental setup.

Because of the numerous geomagnetic regions crossed by the satellite over its $\sim$92-minutes orbit, the proton energy spectrum
was evaluated for sixteen different vertical geomagnetic cutoff intervals, estimated using the satellite position and the St\"{o}rmer approximation. 
The updated (2010) version of the IGRF (\url{https://www.ngdc.noaa.gov/IAGA/vmod/igrf.html}) was used. 
The final fluxes were then evaluated following the approach described in  \cite{Adriani_el_2015}.

It was observed that the high energy part of the resulting spectra had a
systematic time dependence beyond statistical uncertainties with the fluxes varying 
of several percent between 2010 and 2014.
This was corrected following the procedure adopted in  \cite{Adriani_2013}: the fluxes 
were normalized at high energy (30-50 GeV) to the proton flux measured over the period July 2006 - March 2008 (i.e. the proton spectrum of \cite{Adriani_2011} lowered by $3.2\%$ as 
explained in \cite{Adriani_2013}). The uncertainties on these normalization factors, of the order of one percent, were treated as a systematic uncertainty.

\begin{figure}
\centering
\includegraphics[width=.95\textwidth]{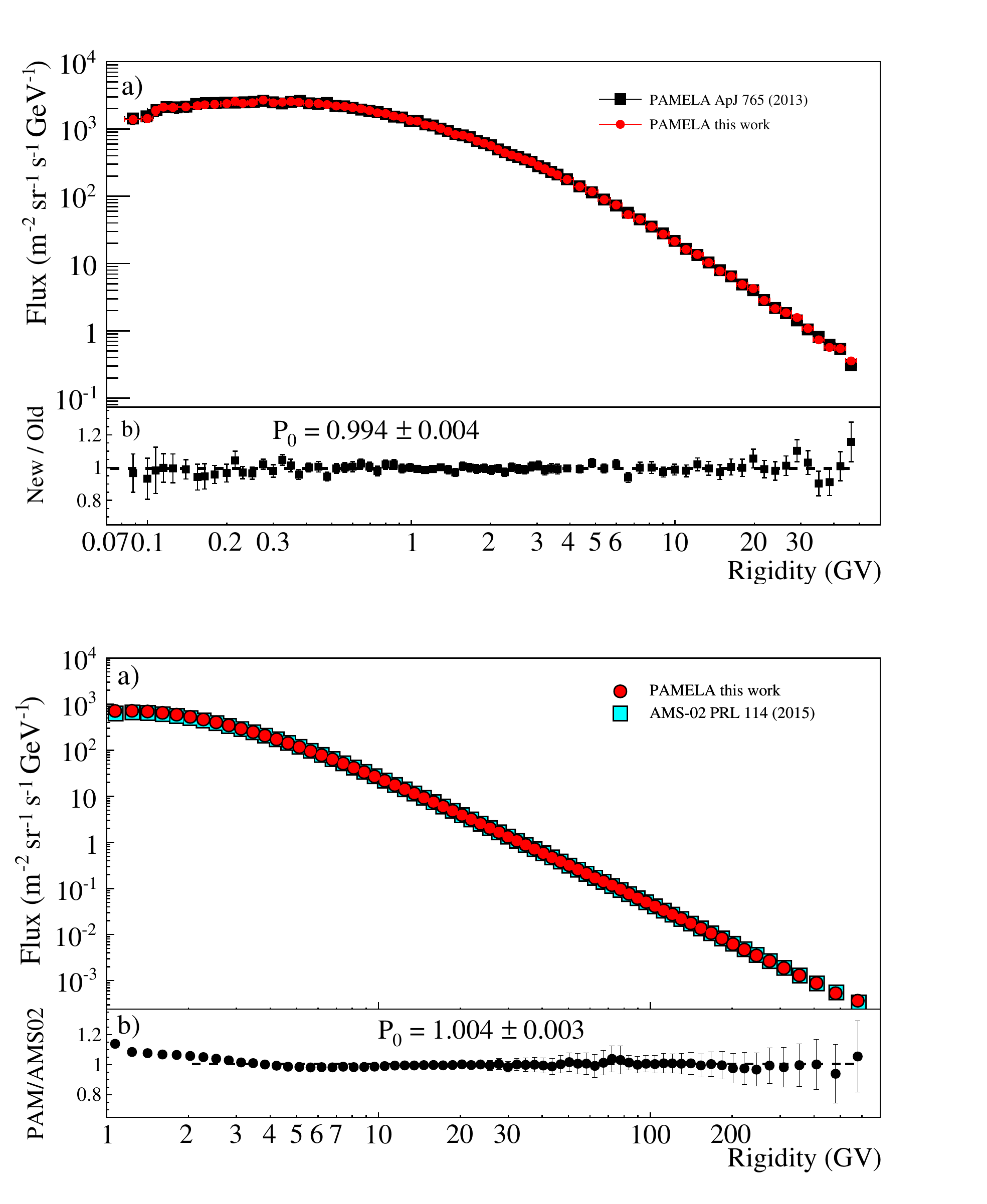}
\caption{Top panel: the proton fluxes measured during the Carrington rotation 2091  
(December 7, 2009 - January 3, 2010) obtained in this analysis (red circles) 
 and the corresponding one obtained in a previous work \citep{Adriani_2013}, (black squares).
 The error bars are statistical and the shaded
area represents the quadratic sum of all systematic uncertainties.  
The ratio between the two results is shown in panel b as well as the value obtained from a constant fit 
(dashed black line) performed on this ratio. Errors are statistical only. Bottom panel: the PAMELA 
fluxes averaged over the period from May 19, 2011  to November 26 2013 compared with concurrent
AMS-02 measurement \citep{AMS02proton} (panel a). Panel b shows 
the ratio between the two measurements.
P$_0$ is the result of a linear fit on this flux ratio above 2 GV. Only statistical errors are showed.}
\label{fig1}
\end{figure}

Other systematic uncertainties were due to the efficiencies evaluation and the unfolding procedure as discussed in \cite{Adriani_2013,Munini_2015,Adriani_el_2015}. 
The total systematic uncertainty shown in Figures \ref{fig1} and \ref{fig2} and in Table \ref{tab1} was obtained quadratically summing the various
systematic errors. 
This systematic uncertainty was about  8$\%$ over the whole energy range and time period. 

Figure \ref{fig1} (top panel a) shows the comparison of the proton fluxes measured during Carrington rotation 2091 (December 7, 2009 - January 3, 2010) obtained in this
analysis with the corresponding ones from \cite{Adriani_2013}. As can be seen from the constant fit performed on the ratio (statistical errors only)
between the two results (top panel b), there is an excellent agreement ($\approx 1\%$) over the whole energy range.
A similar agreement, Figure \ref{fig1} bottom panel a, is also found comparing the PAMELA fluxes averaged over the period from May 19, 2011 to November 26, 
2013 with the corresponding AMS-02 proton 
fluxes \citep{AMS02proton} taken over the same time period. Bottom panel b shows the ratio between the two measurements along with a 
constant fit to the data performed above 2 GV. An excellent agreement can be seen at these rigidities. 
At lower rigidities the PAMELA proton fluxes are systematically higher by about 10$\%$.
This discrepancy could be due to differences in data exclusion periods during solar events and Forbush decreases that have major effects below 2 GV.

\section{Results}

A total of 36 proton energy spectra were obtained from the minimum to the maximum activity of solar cycle 24. 
Data from the Carrington rotations number 2095 to the 2102 are missing because of a system shut-down for satellite maintenance operations.
Moreover, the Carrington rotations number 2113, 2115, 2121, 2123, 2125, 2126, 2135, 2136, 2137, 
2143, 2145 are also missing because of the presence of solar energetic particles as previous explained. 

\begin{figure}
  \centering
      \includegraphics[width=\textwidth]{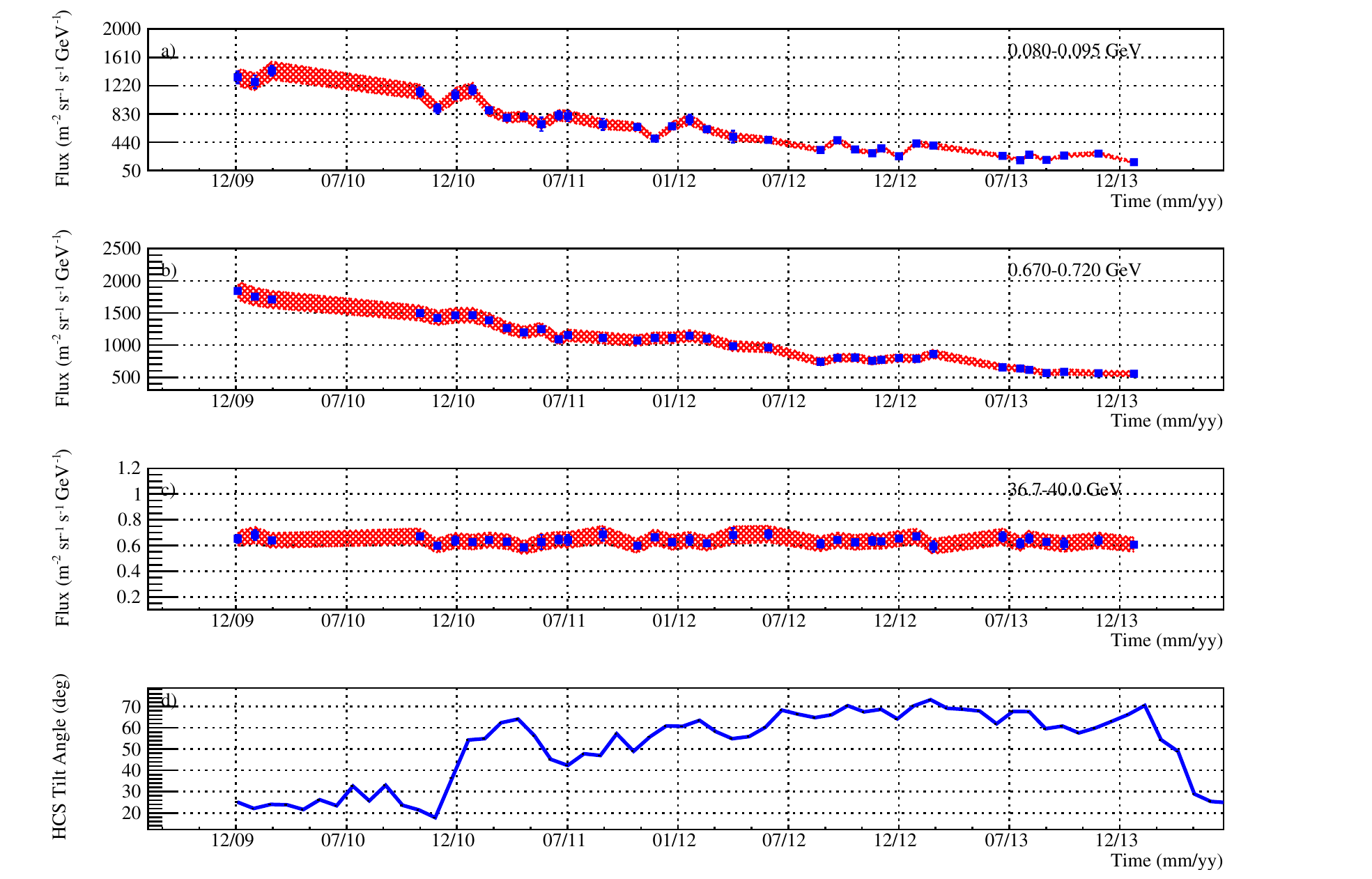}
  \caption{The time profile of the proton flux for three energy intervals: 0.08-0.095 GeV (panel a), 0.67-0.72 GeV (panel b) and 36.7-40 GeV (panel c) along with the time profile 
  of the tilt angle obtained with radial boundary conditions taken from Wilcox Solar Observatory at http://wso.stanford.edu/ (panel d). The error bars indicate the statistical errors while  the shaded
area  the systematic uncertainties.}
\label{fig2}
\end{figure}

Figure \ref{fig2} (panels a,b,c) shows the time profiles of the proton fluxes for three illustrative energy intervals along with the HCS tilt
angle data obtained with radial boundary conditions taken from Wilcox Solar Observatory at http://wso.stanford.edu/.
\footnote{The tilt angle \citep{tilt} represents the misalignment of the magnetic dipole axis of the Sun with respect to the solar
rotational axis and is one of the best proxies for charged particles in cosmic rays because its time variations are related globally to the
solar magnetic field.} (panel d). The red shaded areas represent the  systematic uncertainties while the error bars represent the statistical errors. 
Starting from late 2009, the tilt angle rapidly increased from low values typical for a period of solar minimum activity 
 reaching its maximum value in mid 2012. 
 During this time period the proton fluxes between 0.08-0.095 GeV and between 0.67-0.72 GeV showed a sharp decrease of about a factor 4 and about 
 2.5 respectively, Figure \ref{fig2} panels a and b.
These large values of the tilt angle had basically been maintained, except for a few relatively short periods of decreased values, until 
the end of 2013.   After this period, the tilt angle has decreased systematically, indicating that solar modulation has
turned around to enter a new solar minimum epoch. 
The proton flux after mid 2012 continued to decrease until February 2014 when it reached its minimum intensity. In this time window the fluxes 
between 0.08-0.095 GeV and between 0.67-0.72 GeV decreased by about a factor 2.2 and by about a factor 1.4 respectively, Figure \ref{fig2} panels a and b.
As expected between 36.7-40 GeV, Figure \ref{fig2} panel c, the proton flux is constant with time.

\begin{figure}
  \centering
      \includegraphics[width=\textwidth]{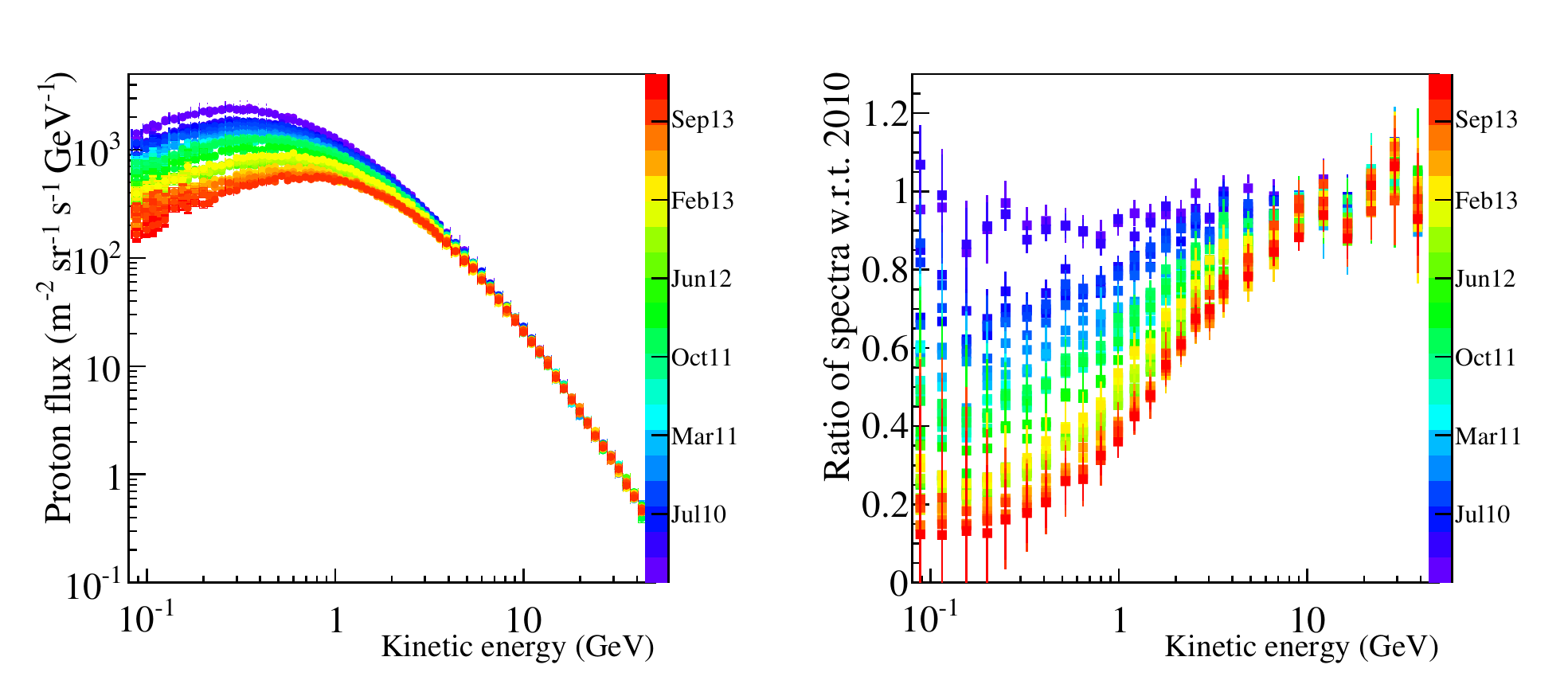}
  \caption{Left panel: the evolution of the proton spectra from the minimum to the maximum activity 
  of solar cycle 24, from January 2010 (blue), to February 2014 (red). Right panel: the variation of the proton 
  differential intensity with respect to the first proton
spectrum of 2010.}
  \label{fig3}
\end{figure}

Figure \ref{fig3} (left panel) shows the totality of the proton fluxes evaluated from January 3, 2010 (blue data points) until February 11, 2014  (red data points) on a 
Carringoton rotation basis.
The right panel of Figure \ref{fig3} shows the variation of the proton intensities with respect to the first Carrington rotation of 2010.
The energy dependence of the solar modulation is particularly evident from this figure: the low energy protons are the most affected with a 
decrease of nearly a factor 10 from the minimum to the maximum solar activity while above $\sim$ 30 GV the proton fluxes do not show any 
temporal variation within the measurement uncertainties. 
Table \ref{tab1} presents the galactic proton spectra measured by the PAMELA experiment over four time periods. 
These data illustrate how the proton spectra evolved from early 2010 to early 2014.
The complete data set can be found at the ASI Space Science Data Center, where all the proton energy spectra are 
retrievable from the Cosmic Ray Data Base (\url{http://tools.asdc.asi.it/CosmicRays/chargedCosmicRays.jsp}).

\section{Data interpretation and discussion}

As shown in Figure \ref{fig3} left panel the observed spectra became progressively harder with increasing solar activity 
as fewer low energy protons were able to reach the Earth. The spectral peaks (turning point in the value of the maximum flux of each spectrum) 
consequently shifted systematically to higher energy values. From 2010 to 2014 the kinetic energy value of the peak shifted from about 350 MeV to 700 MeV.
The adiabatic energy loss signature (spectral shape below the turn-energy proportional to E) became therefore more evident with solar maximum spectra. 
This confirms that adiabatic energy losses for protons (and GCR nuclei) are a significantly important part of the solar
modulation process in the heliosphere (see also \cite{Potgieter_Vos_2017}).

\begin{figure}
  \centering
      \includegraphics[width=\textwidth]{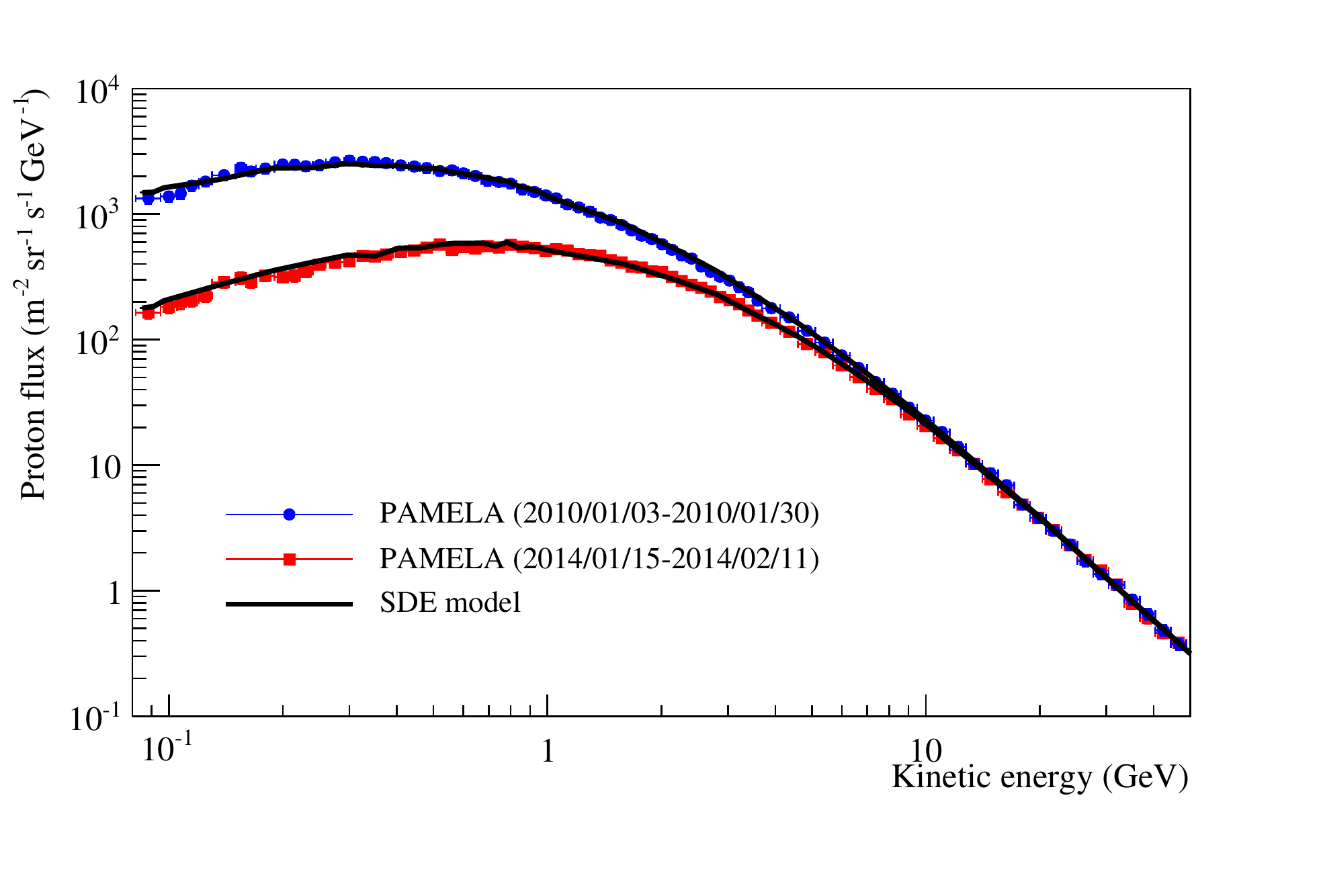}
  \caption{PAMELA proton spectra registered in two Carrington rotations during 2010 (blue circles) and 2014 (red squares).
  Stochastic differential equations (SDE) model (black dashed curve) is superimposed to the observational data.}
  \label{fig4}
\end{figure}

In Figure  \ref{fig4} the PAMELA proton spectra measured in January 2010 and in early 2014 are overlaid with the corresponding computed spectra. 
A full 3D numerical model based on solving Parker’s transport equation \citep{Parker_1965} with the so-called stochastic
differential equation approach was used to compute the proton differential intensity at Earth. 
This modeling approach and its validation against proton observations from PAMELA for the period 2006 to
2009 was published in detail by \cite{Raath_2016}. For a description of a global approach to the modeling of GCR’s in the heliosphere, see also \cite{Potgieter_2017}.

The assumed LIS for protons is chosen according to \cite{Vos_2015}. The modulation volume is assumed to be spherical with the heliopause (HP)
position at 122 AU. The HP is considered to be the outer modulation boundary. For the period 2010 to 2014, changes in the heliospheric current
sheet's tilt angle, as shown in Figure \ref{fig2} panel d, are incorporated into the model together with corresponding changes in the magnitude of 
the solar magnetic field B as observed at the Earth.  These values were averaged for at least the previous 12 months as representation of estimated modulation conditions in the heliosphere
for the prior year. This is based on the average time it  takes for the frozen-in magnetic field and tilt angles to propagate from the Sun to the HP, being carried outward at the average speed of the solar wind.
These averaged values
therefore represent a proxy for the global modulation conditions that prevailed throughout the heliosphere for the time periods considered here.
The three diffusion coefficients in this 3D approach are parallel and perpendicular, in the radial and polar directions, to the global magnetic
field, and are assumed to scale as 1/B, which is the most straight forward approach from a diffusion theory point of view. This follows the basic 
modelling approach described also by \cite{Potgieter_el_2015}. The drift coefficient scales also as 1/B, assuming weak scattering as
explained by \cite{Ngobeni_2015}.

These computed spectra for 2010 and 2014 are shown in Figure \ref{fig4} together with the corresponding observations. Evidently, the model reproduces the features of the two spectra well over this wide energy range, in particular the intensity values where the spectra peak and
how this peak shifts to higher energies while the spectrum decreases with increased modulation. Reproducing the 2010 spectrum (during an A$<$0 magnetic polarity cycle\footnote{In the Sun magnetic field the dipole term nearly always dominates the magnetic field of the solar wind. A is defined as the projection  of this dipole on the solar rotation axis.}) required relatively minor changes to the modulation parameters used by \cite{Raath_2016}. However, in order
to reproduce the 2014 spectrum (during an A$>$0 magnetic polarity cycle), with the amount of modulation additionally occurring as shown in Figure \ref{fig3}, the diffusion coefficients had to be decreased by a factor of 2 with respect to the 2010 values. Simultaneously, the drift coefficient had to be reduced to only 10\% of the solar minimum value. This illustrates that reproducing the total amount of modulation occurring from maximum GCR intensity in
early 2010 to minimum intensity in 2014 requires about a factor of 2 increase in the effectiveness of diffusion while drifts had to be significantly reduced, otherwise the intensity levels would have remained far too high with increasing solar modulation for this A$>$0
magnetic cycle. 

\section{Conclusion}

The observations presented here illustrate the total modulation that had occurred from minimum modulation (highest intensity) of GCRs 
to maximum modulation (lowest intensity) under a relatively quiet Sun and subsequently also the heliosphere. This provides a unique opportunity to study the modulation of GCRs under such extraordinary conditions. In particular, combined with the observed electron to positron ratios 
reported by PAMELA in \cite{Adriani_2016}, it provides information useful to understand how diffusion and drifts effects vary with time and energy.

\newpage

\begin{longrotatetable}
\begin{deluxetable}{lcccc}
\tablecaption{Proton flux measured by PAMELA over four time periods. The first and second errors represent the statistical
and systematic uncertainties, respectively.\label{tab1}}
\tablewidth{700pt}
\tabletypesize{\scriptsize}
\tablehead{
\colhead{Kinetic Energy (GeV)} &                                 & \colhead{Flux{(m$^2$ s sr GeV)$^{-1}$}} &                                 &                                 \\
                               & \colhead{2010/01/03-2010/01/30} & \colhead{2011/04/12-2011/05/09}         & \colhead{2012/08/15-2012/09/11} & \colhead{2014/01/15-2014/02/15} \\
} 
\startdata
 0.082-0.095 & 1334.967 $\pm$ 82.677 $\pm$ 127.149 & 795.308 $\pm$ 47.742 $\pm$ 76.072 & 333.947 $\pm$ 40.400 $\pm$ 36.724 & 164.583 $\pm$ 25.663 $\pm$ 18.902 \\ 
0.095-0.105 & 1382.404 $\pm$ 96.083 $\pm$ 133.199 & 836.770 $\pm$ 55.939 $\pm$ 81.172 & 415.946 $\pm$ 51.653 $\pm$ 45.755 & 181.962 $\pm$ 30.565 $\pm$ 21.320 \\ 
0.105-0.110 & 1462.203 $\pm$ 138.301 $\pm$ 148.366 & 939.265 $\pm$ 83.015 $\pm$ 95.267 & 345.066 $\pm$ 65.980 $\pm$ 43.123 & 199.548 $\pm$ 44.385 $\pm$ 25.410 \\ 
0.110-0.120 & 1682.318 $\pm$ 103.829 $\pm$ 157.830 & 979.690 $\pm$ 59.402 $\pm$ 92.551 & 438.529 $\pm$ 51.925 $\pm$ 46.809 & 205.254 $\pm$ 31.346 $\pm$ 22.790 \\ 
0.120-0.130 & 1816.430 $\pm$ 106.984 $\pm$ 149.757 & 963.679 $\pm$ 58.432 $\pm$ 81.848 & 494.574 $\pm$ 54.318 $\pm$ 47.484 & 223.106 $\pm$ 32.338 $\pm$ 23.175 \\ 
0.130-0.150 & 2040.187 $\pm$ 79.696 $\pm$ 180.845 & 1075.864 $\pm$ 43.351 $\pm$ 96.031 & 552.629 $\pm$ 40.069 $\pm$ 52.988 & 286.300 $\pm$ 25.680 $\pm$ 28.578 \\ 
0.150-0.160 & 2343.113 $\pm$ 119.948 $\pm$ 214.642 & 1046.064 $\pm$ 59.954 $\pm$ 96.757 & 542.123 $\pm$ 55.489 $\pm$ 55.293 & 309.547 $\pm$ 37.346 $\pm$ 32.766 \\ 
0.160-0.170 & 2185.130 $\pm$ 114.921 $\pm$ 199.299 & 1151.039 $\pm$ 62.378 $\pm$ 105.827 & 602.654 $\pm$ 58.035 $\pm$ 60.533 & 287.579 $\pm$ 35.579 $\pm$ 30.646 \\ 
0.170-0.190 & 2302.665 $\pm$ 82.801 $\pm$ 201.218 & 1179.811 $\pm$ 44.381 $\pm$ 103.910 & 625.510 $\pm$ 41.630 $\pm$ 58.856 & 322.769 $\pm$ 26.411 $\pm$ 31.053 \\ 
0.190-0.210 & 2484.283 $\pm$ 85.461 $\pm$ 215.673 & 1304.952 $\pm$ 46.483 $\pm$ 114.144 & 668.770 $\pm$ 42.942 $\pm$ 62.440 & 314.053 $\pm$ 25.881 $\pm$ 30.283 \\ 
0.210-0.220 & 2469.407 $\pm$ 82.545 $\pm$ 220.425 & 1282.348 $\pm$ 44.210 $\pm$ 115.907 & 635.254 $\pm$ 39.659 $\pm$ 61.927 & 322.497 $\pm$ 24.950 $\pm$ 32.828 \\ 
0.220-0.240 & 2404.784 $\pm$ 57.377 $\pm$ 207.994 & 1323.680 $\pm$ 31.696 $\pm$ 115.256 & 643.178 $\pm$ 28.150 $\pm$ 59.086 & 347.635 $\pm$ 18.215 $\pm$ 32.578 \\ 
0.240-0.260 & 2452.868 $\pm$ 57.790 $\pm$ 211.535 & 1381.816 $\pm$ 32.343 $\pm$ 119.807 & 703.898 $\pm$ 29.401 $\pm$ 64.278 & 395.317 $\pm$ 19.325 $\pm$ 36.942 \\ 
0.260-0.290 & 2587.382 $\pm$ 48.354 $\pm$ 219.281 & 1382.849 $\pm$ 26.389 $\pm$ 118.183 & 744.302 $\pm$ 24.665 $\pm$ 66.175 & 411.836 $\pm$ 16.033 $\pm$ 37.071 \\ 
0.290-0.310 & 2660.621 $\pm$ 59.944 $\pm$ 228.362 & 1411.766 $\pm$ 32.628 $\pm$ 122.150 & 742.029 $\pm$ 30.165 $\pm$ 67.279 & 420.169 $\pm$ 19.762 $\pm$ 38.487 \\ 
0.310-0.340 & 2614.418 $\pm$ 48.414 $\pm$ 221.109 & 1405.834 $\pm$ 26.549 $\pm$ 119.878 & 734.236 $\pm$ 24.507 $\pm$ 65.126 & 465.419 $\pm$ 16.922 $\pm$ 41.386 \\ 
0.340-0.360 & 2600.506 $\pm$ 59.064 $\pm$ 222.681 & 1427.089 $\pm$ 32.739 $\pm$ 123.237 & 740.257 $\pm$ 30.166 $\pm$ 66.846 & 460.405 $\pm$ 20.562 $\pm$ 41.725 \\ 
0.360-0.390 & 2549.032 $\pm$ 47.743 $\pm$ 215.811 & 1412.139 $\pm$ 26.604 $\pm$ 120.255 & 774.639 $\pm$ 25.225 $\pm$ 68.524 & 476.739 $\pm$ 17.072 $\pm$ 41.993 \\ 
0.390-0.430 & 2443.305 $\pm$ 40.470 $\pm$ 205.093 & 1414.240 $\pm$ 23.069 $\pm$ 119.391 & 751.346 $\pm$ 21.516 $\pm$ 65.442 & 500.676 $\pm$ 15.152 $\pm$ 43.648 \\ 
0.430-0.460 & 2393.106 $\pm$ 46.216 $\pm$ 202.300 & 1386.433 $\pm$ 26.386 $\pm$ 118.106 & 779.713 $\pm$ 25.300 $\pm$ 68.331 & 512.399 $\pm$ 17.702 $\pm$ 45.058 \\ 
0.460-0.500 & 2328.825 $\pm$ 39.511 $\pm$ 195.569 & 1426.760 $\pm$ 23.225 $\pm$ 120.411 & 770.670 $\pm$ 21.819 $\pm$ 66.803 & 540.540 $\pm$ 15.766 $\pm$ 46.855 \\ 
0.500-0.540 & 2201.917 $\pm$ 38.488 $\pm$ 185.090 & 1358.087 $\pm$ 22.725 $\pm$ 114.736 & 773.471 $\pm$ 21.932 $\pm$ 67.123 & 571.107 $\pm$ 16.240 $\pm$ 49.419 \\ 
0.540-0.580 & 2232.442 $\pm$ 38.796 $\pm$ 187.522 & 1312.368 $\pm$ 22.390 $\pm$ 110.961 & 774.749 $\pm$ 22.017 $\pm$ 67.089 & 518.287 $\pm$ 15.498 $\pm$ 44.889 \\ 
0.580-0.620 & 2110.885 $\pm$ 37.741 $\pm$ 177.592 & 1308.983 $\pm$ 22.407 $\pm$ 110.699 & 759.016 $\pm$ 21.854 $\pm$ 65.851 & 546.131 $\pm$ 15.941 $\pm$ 47.238 \\ 
0.620-0.670 & 2013.549 $\pm$ 32.989 $\pm$ 168.502 & 1272.052 $\pm$ 19.810 $\pm$ 107.028 & 776.185 $\pm$ 19.833 $\pm$ 66.748 & 532.933 $\pm$ 14.124 $\pm$ 45.789 \\ 
0.670-0.720 & 1843.377 $\pm$ 31.596 $\pm$ 154.484 & 1198.768 $\pm$ 19.288 $\pm$ 101.059 & 739.141 $\pm$ 19.425 $\pm$ 63.661 & 556.023 $\pm$ 14.472 $\pm$ 47.701 \\ 
0.720-0.770 & 1802.870 $\pm$ 31.308 $\pm$ 151.032 & 1124.218 $\pm$ 18.736 $\pm$ 94.805 & 718.388 $\pm$ 19.226 $\pm$ 61.941 & 542.074 $\pm$ 14.337 $\pm$ 46.560 \\ 
0.770-0.830 & 1749.204 $\pm$ 28.240 $\pm$ 146.298 & 1121.779 $\pm$ 17.137 $\pm$ 94.240 & 719.320 $\pm$ 17.639 $\pm$ 61.656 & 567.905 $\pm$ 13.440 $\pm$ 48.460 \\ 
0.830-0.890 & 1570.682 $\pm$ 26.852 $\pm$ 131.498 & 1087.582 $\pm$ 16.923 $\pm$ 91.463 & 711.780 $\pm$ 17.622 $\pm$ 61.110 & 548.560 $\pm$ 13.250 $\pm$ 46.785 \\ 
0.890-0.960 & 1500.112 $\pm$ 24.349 $\pm$ 125.310 & 1038.532 $\pm$ 15.345 $\pm$ 87.093 & 663.605 $\pm$ 15.808 $\pm$ 56.763 & 537.129 $\pm$ 12.168 $\pm$ 45.622 \\ 
0.960-1.020 & 1409.334 $\pm$ 19.521 $\pm$ 118.330 & 936.771 $\pm$ 12.030 $\pm$ 79.009 & 646.067 $\pm$ 12.807 $\pm$ 55.601 & 507.792 $\pm$ 9.730 $\pm$ 43.461 \\ 
1.020-1.090 & 1335.055 $\pm$ 17.609 $\pm$ 111.752 & 897.745 $\pm$ 10.933 $\pm$ 75.526 & 622.088 $\pm$ 11.675 $\pm$ 53.337 & 525.043 $\pm$ 9.184 $\pm$ 44.684 \\ 
1.090-1.170 & 1194.435 $\pm$ 15.600 $\pm$ 99.908 & 860.732 $\pm$ 10.040 $\pm$ 72.209 & 600.630 $\pm$ 10.772 $\pm$ 51.347 & 511.695 $\pm$ 8.505 $\pm$ 43.441 \\ 
1.170-1.250 & 1130.494 $\pm$ 15.201 $\pm$ 94.684 & 789.504 $\pm$ 9.640 $\pm$ 66.367 & 561.511 $\pm$ 10.457 $\pm$ 48.076 & 480.355 $\pm$ 8.265 $\pm$ 40.828 \\ 
1.250-1.340 & 1043.878 $\pm$ 13.799 $\pm$ 87.331 & 726.467 $\pm$ 8.740 $\pm$ 61.020 & 548.761 $\pm$ 9.786 $\pm$ 46.897 & 469.432 $\pm$ 7.728 $\pm$ 39.799 \\ 
1.340-1.420 & 938.592 $\pm$ 13.907 $\pm$ 78.904 & 721.208 $\pm$ 9.258 $\pm$ 60.809 & 496.622 $\pm$ 9.909 $\pm$ 42.723 & 465.878 $\pm$ 8.191 $\pm$ 39.672 \\ 
1.420-1.520 & 895.874 $\pm$ 12.176 $\pm$ 75.005 & 662.716 $\pm$ 7.953 $\pm$ 55.659 & 488.957 $\pm$ 8.819 $\pm$ 41.800 & 428.841 $\pm$ 7.047 $\pm$ 36.392 \\ 
1.520-1.620 & 818.801 $\pm$ 11.657 $\pm$ 68.713 & 606.027 $\pm$ 7.616 $\pm$ 51.005 & 445.967 $\pm$ 8.440 $\pm$ 38.274 & 411.672 $\pm$ 6.917 $\pm$ 34.947 \\ 
1.620-1.720 & 741.816 $\pm$ 11.099 $\pm$ 62.396 & 562.603 $\pm$ 7.341 $\pm$ 47.396 & 435.086 $\pm$ 8.343 $\pm$ 37.298 & 382.659 $\pm$ 6.672 $\pm$ 32.581 \\ 
1.720-1.830 & 673.281 $\pm$ 10.082 $\pm$ 56.557 & 525.954 $\pm$ 6.769 $\pm$ 44.300 & 391.637 $\pm$ 7.549 $\pm$ 33.642 & 375.139 $\pm$ 6.299 $\pm$ 31.831 \\ 
1.830-1.950 & 631.863 $\pm$ 9.364 $\pm$ 53.079 & 490.276 $\pm$ 6.265 $\pm$ 41.286 & 370.506 $\pm$ 7.034 $\pm$ 31.736 & 348.823 $\pm$ 5.823 $\pm$ 29.618 \\ 
1.950-2.070 & 575.142 $\pm$ 8.955 $\pm$ 48.483 & 431.415 $\pm$ 5.888 $\pm$ 36.422 & 340.164 $\pm$ 6.748 $\pm$ 29.239 & 346.174 $\pm$ 5.814 $\pm$ 29.424 \\ 
2.070-2.200 & 519.375 $\pm$ 7.169 $\pm$ 43.745 & 415.325 $\pm$ 4.927 $\pm$ 35.038 & 332.828 $\pm$ 5.681 $\pm$ 28.518 & 316.716 $\pm$ 4.726 $\pm$ 26.912 \\ 
2.200-2.330 & 468.252 $\pm$ 6.822 $\pm$ 39.547 & 381.339 $\pm$ 4.730 $\pm$ 32.242 & 314.474 $\pm$ 5.533 $\pm$ 27.009 & 293.339 $\pm$ 4.559 $\pm$ 24.986 \\ 
2.330-2.480 & 442.058 $\pm$ 6.183 $\pm$ 37.266 & 338.366 $\pm$ 4.155 $\pm$ 28.567 & 290.548 $\pm$ 4.963 $\pm$ 24.913 & 272.463 $\pm$ 4.097 $\pm$ 23.155 \\ 
2.480-2.620 & 382.832 $\pm$ 5.963 $\pm$ 32.457 & 314.857 $\pm$ 4.152 $\pm$ 26.688 & 271.590 $\pm$ 4.974 $\pm$ 23.412 & 258.542 $\pm$ 4.132 $\pm$ 22.078 \\ 
2.620-2.780 & 346.394 $\pm$ 5.307 $\pm$ 29.322 & 294.360 $\pm$ 3.757 $\pm$ 24.897 & 250.850 $\pm$ 4.477 $\pm$ 21.578 & 241.663 $\pm$ 3.736 $\pm$ 20.586 \\ 
2.780-2.940 & 317.209 $\pm$ 5.076 $\pm$ 26.883 & 264.513 $\pm$ 3.561 $\pm$ 22.461 & 223.179 $\pm$ 4.225 $\pm$ 19.268 & 218.276 $\pm$ 3.549 $\pm$ 18.634 \\ 
2.940-3.120 & 295.253 $\pm$ 4.613 $\pm$ 25.014 & 241.820 $\pm$ 3.209 $\pm$ 20.506 & 206.612 $\pm$ 3.834 $\pm$ 17.809 & 206.426 $\pm$ 3.255 $\pm$ 17.594 \\ 
3.120-3.300 & 262.302 $\pm$ 4.347 $\pm$ 22.278 & 216.080 $\pm$ 3.034 $\pm$ 18.367 & 188.554 $\pm$ 3.667 $\pm$ 16.281 & 191.244 $\pm$ 3.136 $\pm$ 16.322 \\ 
3.300-3.490 & 238.599 $\pm$ 4.036 $\pm$ 20.280 & 201.418 $\pm$ 2.851 $\pm$ 17.135 & 179.749 $\pm$ 3.487 $\pm$ 15.538 & 170.347 $\pm$ 2.884 $\pm$ 14.573 \\ 
3.490-3.690 & 204.508 $\pm$ 3.642 $\pm$ 17.442 & 185.956 $\pm$ 2.670 $\pm$ 15.843 & 162.815 $\pm$ 3.235 $\pm$ 14.113 & 155.731 $\pm$ 2.689 $\pm$ 13.353 \\ 
3.690-4.120 & 178.114 $\pm$ 2.322 $\pm$ 14.930 & 153.363 $\pm$ 1.656 $\pm$ 12.850 & 139.847 $\pm$ 2.047 $\pm$ 11.850 & 136.054 $\pm$ 1.717 $\pm$ 11.440 \\ 
4.120-4.590 & 150.564 $\pm$ 2.046 $\pm$ 12.646 & 128.540 $\pm$ 1.453 $\pm$ 10.794 & 114.930 $\pm$ 1.778 $\pm$ 9.773 & 115.714 $\pm$ 1.517 $\pm$ 9.753 \\ 
4.590-5.110 & 117.786 $\pm$ 1.723 $\pm$ 9.924 & 103.522 $\pm$ 1.241 $\pm$ 8.711 & 95.778 $\pm$ 1.544 $\pm$ 8.163 & 92.180 $\pm$ 1.288 $\pm$ 7.787 \\ 
5.110-5.680 & 94.763 $\pm$ 1.475 $\pm$ 8.008 & 85.760 $\pm$ 1.079 $\pm$ 7.239 & 80.949 $\pm$ 1.355 $\pm$ 6.919 & 79.408 $\pm$ 1.141 $\pm$ 6.725 \\ 
5.680-6.300 & 74.839 $\pm$ 1.255 $\pm$ 6.340 & 66.203 $\pm$ 0.909 $\pm$ 5.605 & 64.084 $\pm$ 1.155 $\pm$ 5.497 & 62.411 $\pm$ 0.968 $\pm$ 5.300 \\ 
6.300-6.990 & 59.473 $\pm$ 1.059 $\pm$ 5.048 & 53.257 $\pm$ 0.773 $\pm$ 4.516 & 50.253 $\pm$ 0.970 $\pm$ 4.319 & 50.251 $\pm$ 0.824 $\pm$ 4.273 \\ 
6.990-7.740 & 45.962 $\pm$ 0.893 $\pm$ 3.925 & 43.370 $\pm$ 0.670 $\pm$ 3.693 & 42.767 $\pm$ 0.859 $\pm$ 3.688 & 40.426 $\pm$ 0.710 $\pm$ 3.455 \\ 
7.740-8.570 & 37.064 $\pm$ 0.763 $\pm$ 3.170 & 35.503 $\pm$ 0.577 $\pm$ 3.030 & 32.609 $\pm$ 0.713 $\pm$ 2.826 & 33.508 $\pm$ 0.616 $\pm$ 2.868 \\ 
8.570-9.480 & 28.753 $\pm$ 0.512 $\pm$ 2.471 & 25.887 $\pm$ 0.379 $\pm$ 2.222 & 26.297 $\pm$ 0.492 $\pm$ 2.286 & 25.380 $\pm$ 0.408 $\pm$ 2.184 \\ 
9.480-10.480 & 22.697 $\pm$ 0.435 $\pm$ 1.956 & 21.708 $\pm$ 0.331 $\pm$ 1.867 & 21.052 $\pm$ 0.420 $\pm$ 1.840 & 20.517 $\pm$ 0.350 $\pm$ 1.773 \\ 
10.480-11.570 & 18.376 $\pm$ 0.375 $\pm$ 1.594 & 17.469 $\pm$ 0.284 $\pm$ 1.512 & 17.000 $\pm$ 0.361 $\pm$ 1.493 & 16.431 $\pm$ 0.300 $\pm$ 1.426 \\ 
11.570-12.770 & 13.985 $\pm$ 0.312 $\pm$ 1.218 & 13.673 $\pm$ 0.239 $\pm$ 1.187 & 13.349 $\pm$ 0.304 $\pm$ 1.177 & 13.135 $\pm$ 0.256 $\pm$ 1.145 \\ 
12.770-14.090 & 10.229 $\pm$ 0.255 $\pm$ 0.899 & 10.112 $\pm$ 0.196 $\pm$ 0.884 & 10.645 $\pm$ 0.259 $\pm$ 0.944 & 10.221 $\pm$ 0.216 $\pm$ 0.896 \\ 
14.090-15.540 & 8.623 $\pm$ 0.223 $\pm$ 0.761 & 8.455 $\pm$ 0.171 $\pm$ 0.741 & 8.362 $\pm$ 0.219 $\pm$ 0.747 & 7.718 $\pm$ 0.179 $\pm$ 0.680 \\ 
15.540-17.120 & 6.927 $\pm$ 0.192 $\pm$ 0.613 & 6.545 $\pm$ 0.144 $\pm$ 0.577 & 6.278 $\pm$ 0.182 $\pm$ 0.564 & 6.095 $\pm$ 0.152 $\pm$ 0.540 \\ 
17.120-18.860 & 4.854 $\pm$ 0.153 $\pm$ 0.432 & 5.109 $\pm$ 0.121 $\pm$ 0.453 & 5.067 $\pm$ 0.156 $\pm$ 0.458 & 4.840 $\pm$ 0.129 $\pm$ 0.432 \\ 
18.860-20.760 & 3.811 $\pm$ 0.129 $\pm$ 0.343 & 3.756 $\pm$ 0.099 $\pm$ 0.337 & 3.958 $\pm$ 0.132 $\pm$ 0.361 & 3.805 $\pm$ 0.110 $\pm$ 0.342 \\ 
20.760-22.850 & 2.998 $\pm$ 0.091 $\pm$ 0.272 & 2.950 $\pm$ 0.072 $\pm$ 0.266 & 2.933 $\pm$ 0.093 $\pm$ 0.271 & 3.044 $\pm$ 0.080 $\pm$ 0.276 \\ 
22.850-25.150 & 2.332 $\pm$ 0.076 $\pm$ 0.214 & 2.318 $\pm$ 0.061 $\pm$ 0.211 & 2.209 $\pm$ 0.077 $\pm$ 0.208 & 2.307 $\pm$ 0.066 $\pm$ 0.212 \\ 
25.150-27.660 & 1.712 $\pm$ 0.063 $\pm$ 0.159 & 1.770 $\pm$ 0.051 $\pm$ 0.163 & 1.741 $\pm$ 0.065 $\pm$ 0.165 & 1.753 $\pm$ 0.055 $\pm$ 0.162 \\ 
27.660-30.420 & 1.358 $\pm$ 0.053 $\pm$ 0.128 & 1.497 $\pm$ 0.045 $\pm$ 0.139 & 1.448 $\pm$ 0.057 $\pm$ 0.138 & 1.445 $\pm$ 0.048 $\pm$ 0.135 \\ 
30.420-33.440 & 1.111 $\pm$ 0.046 $\pm$ 0.105 & 1.149 $\pm$ 0.038 $\pm$ 0.108 & 1.141 $\pm$ 0.048 $\pm$ 0.111 & 1.116 $\pm$ 0.040 $\pm$ 0.105 \\ 
33.440-36.750 & 0.850 $\pm$ 0.038 $\pm$ 0.082 & 0.840 $\pm$ 0.031 $\pm$ 0.080 & 0.799 $\pm$ 0.039 $\pm$ 0.080 & 0.787 $\pm$ 0.032 $\pm$ 0.075 \\ 
36.750-40.390 & 0.653 $\pm$ 0.032 $\pm$ 0.064 & 0.585 $\pm$ 0.024 $\pm$ 0.057 & 0.613 $\pm$ 0.032 $\pm$ 0.062 & 0.607 $\pm$ 0.027 $\pm$ 0.060 \\ 
40.390-44.370 & 0.485 $\pm$ 0.026 $\pm$ 0.048 & 0.465 $\pm$ 0.021 $\pm$ 0.046 & 0.442 $\pm$ 0.026 $\pm$ 0.046 & 0.460 $\pm$ 0.022 $\pm$ 0.046 \\ 
44.370-48.740 & 0.376 $\pm$ 0.022 $\pm$ 0.038 & 0.389 $\pm$ 0.018 $\pm$ 0.039 & 0.368 $\pm$ 0.023 $\pm$ 0.038 & 0.386 $\pm$ 0.019 $\pm$ 0.039 \\
\enddata
\end{deluxetable}
\end{longrotatetable}

\end{document}